# A bayesian reanalysis of the phase III aducanumab (ADU) trial

Tommaso Costa[1,2,3], Franco Cauda[1,2,3]

[1]GCS-fMRI, Koelliker Hospital and Department of Psychology, University of Turin, Turin, Italy
[2]Department of Psychology, University of Turin, Turin, Italy
[3]FOCUS Lab, Department of Psychology, university of Turin, Turin, Italy.

Short title: Bayesian reanalysis of ADU trial

Correspondence

Tommaso Costa, Department of Psychology, Via Verdi 10, 10124 Turin, Italy

E-mail: tommaso.costa@unito.it

## Abstract

In this article we have conducted a reanalysis of the phase III aducanumab (ADU) summary statistics announced by Biogen, in particular the result of the Clinical Dementia Rating-Sum of Boxes (CDR-SB). The results showed that the evidence on the efficacy of the drug is very low and a more clearer view of the results of clinical trials are presented in the Bayesian framework that can be useful for future development and research in the field.

# Introduction

In March 2019 Biogen announced the stop their phase III trial of the drug Aducanumab (ADU) due to futility. Seven months later, in December 2019, Biogen claimed for efficacy of ADU. The results were presented at an international meeting in San Diego, California, and the slide with the results were released on line ([https://investors.biogen.com/static-files/ddd45672-9c7e-4c99-8a06-3b557697c06f)](https://investors.biogen.com/static-files/ddd45672-9c7e-4c99-8a06-3b557697c06f), for a deeply review of the trials and clinical consideration see Knopman et al. (2021). The claim was of evidence in the efficacy for high dose ADU in the halted trials. On the base of this last evidence the Biogen submitted a New Drug Application to the Food and Drug Administration (FDA) in July 2020. However, questions were quickly raised regarding the study's and many doubts have been raised by different researchers including the Office of Biostatistics within the Office of Translational Sciences (OTS) of the FDA (Food and Drugs Administration) declared that substantial evidence of effectiveness had not been provided, see the following link for all the information furnished by the FDA ([https://www.accessdata.fda.gov/drugsatfda_docs/nda/2021/761178_Orig1s000TOC.cfm)](https://www.accessdata.fda.gov/drugsatfda_docs/nda/2021/761178_Orig1s000TOC.cfm).

The Food and Drug Administration (FDA) uses a $p < 0.05$ null-hypothesis significance testing framework (NHST) to evaluate "substantial evidence" for drug efficacy.

The NHST framework, is associated with a number of problems (Goodman, 1999). First, the $p$-value is prone to misinterpretation, leading to overestimate the evidence against the null-hypothesis. Second, the criterion p<0.05 induces a binary "all or none" used as a reference to either accept or reject the null hypothesis. Consequently, the current state of affairs harbors the following dangers: (a) the FDA may approve a drug whose efficacy is only minimally supported by the data; (b) the evidence in favor of efficacy cannot be assessed on a gradual scale. On the other hand, Bayesian statistics allows for a direct evaluation of the evidence for and against competing hypotheses, not possible in the frequentist method, and also it is possible to assess how strong is the evidence of a treatment effect between the hypothesis.

In light of these questions, we reanalyzed the data furnished, in particular the result of the Clinical Dementia Rating-Sum of Boxes (CDR-SB) that showed a significant $p$ values for the high dose condition in the EMERGE trial, by performing a Bayesian analysis.

# Materials and methods

## Data

The results of the clinical trials are reported at the link [https://investors.biogen.com/static-files/ddd45672-9c7e-4c99-8a06-3b557697c06f](https://investors.biogen.com/static-files/ddd45672-9c7e-4c99-8a06-3b557697c06f). Raw data was not available and the slides at the previous cited link is the only source of data and results available about the phase III trial. Specifically, we analyzed the Clinical Dementia Rating-Sum of Boxes (CDR-SB) available for the EMERGE final data set at week 78 and the ENGAGE final data set. See table 1 for a summary of the results used.

|  | **Low dose** | **High dose** |
|---|---|---|
| EMERGE | $n = 543$<br>$p = 0.09$ | $n = 547$<br>$p = 0.012$ |
| ENGAGE | $n = 547$<br>$p = 0.24$ | $n = 555$<br>$p = 0.82$ |

**Table 1**. The row represents the public data of the EMERGE and the ENGAGE trials. Here $n$ is the sample size, $p$ the original $p$ values.

## Statistical Method

To alleviate the number of problem of the NHST framework described in the introduction we used a Bayesian framework. In particular, we used the Bayes Factor (BF), which, in its simplest form is also called likelihood ratio, is a comparison of how well two hypotheses predict the data. The hypothesis that predicts the observed data better is the one that is said to have more evidence supporting it. The equation of the BF is:

$$BF_{01} = \frac{L(Data|given\ the\ null\ hypothesis)}{L(Data|given\ the\ alternative\ hypothesis)}$$

where $L$ is the likelihood of the null and alternative hypothesis. This ratio quantifies how data support the $H_0$ hypothesis over $H_1$.

The Bayes factor differs in many ways from a NHST framework. First, the Bayes factor is a ratio of probabilities, and it can vary from zero to infinity. It requires two hypotheses, making it clear that for evidence to be against the null hypothesis, it must be for some alternative. Second, the Bayes factor depends on the probability of the observed data alone, not unobserved "long run" results as used for the $p$ value calculation. Thus, factors unrelated to the data that affect the $p$ value, such as stopping rule, do not affect the Bayes factor.

In this case, we have only summary statistics and to calculate the BF we need of a $t$-statistics to calculate the BF. Using the sample size $N$ and the $p$-value in each condition we can obtain the corresponding $t$-values by the inverse cumulative distribution function of the Student's evaluated at the probability values $p$ using the corresponding degrees of freedom $v = N - 1$.

From the $t$-statistics it possible to obtain a BF, given the null (no effect) and the alternative hypotheses (there is an effect) see Rouder et al. (2009), given by:

$$B_{01} = \left(1 + \frac{t^2}{v}\right)^{-\frac{(v+1)}{2}} \Big/ \int_0^\infty (1 + N_0 g)^{-\frac{1}{2}} \left(1 + \frac{t^2}{(1 + N_0 g)v}\right)^{-\frac{v+1}{2}} (2\pi)^{-\frac{1}{2}} g^{-\frac{3}{2}} e^{-\frac{1}{2g}} dg$$

where $v = n1 + n2 - 2$ is the degree of freedom ans $N_0 = \frac{n_1 n_2}{n_1 + n_2}$ is the effective sample size.

All references to strength of evidence refer to standard conventions for the evidentiary support of Bayes factors (BF) such that 1–3 is classed as anecdotal, 3–10 as moderate, 10–30 as strong, and 30–100 as very strong (Jeffreys, 1961). For Bayes factors below 1, the reciprocal can be taken to obtain the strength of evidence in the opposite direction.

A final analysis is conducted using the result of the EMERGE and ENGAGE trial by a meta-analysis. The idea is that if there are replicate experiments, it seems reasonable, from the Bayesian point of view, that the posterior odds from the first can serve as the prior for the second, and so on. For this analysis, we used a meta-analytic extension of the Bayes factor proposed by Rouder and Morey (2011) defined as:

$$B_{01} = \frac{\prod_{j=1}^{M} g(t_j, N_j - 1, 0)}{\int_0^\infty (\prod_{j=1}^{M} g(t_j, N_j - 1, \delta\sqrt{N_j}) f(\delta) d\delta}$$

where $g$ id the probability density function of the noncentral $t$ and $f$ is the probability density function of the Cauchy.

# Results

All Bayesian statistical analyses of the data were performed in JASP (version 0.92, jasp-stats.org) and using the BayesFactor package in R (Morey and Rouder, 2015) for the meta-analysis.
In table 2 and fig. 1 are show the results of the BF analysis of the two trials.

|        | Low dose              | High dose             |
| ------ | --------------------- | --------------------- |
| EMERGE | $t = 1.69$<br>$BF_{10} = 0.27$ | $t = 2.52$<br>$BF_{10} = 1.54$ |
| ENGAGE | $t = 1.17$<br>$BF_{10} = 0.13$ | $t = 0.23$<br>$BF_{10} = 0.07$ |

**Table 2**. The row represents the results of the bayesian reanalysis of EMERGE and the ENGAGE trials, respectively. Here $t$ is the $t$-values obtained from the $p$-values and $BF_{10}$ is the Bayes Factor of the alternative hypothesis respect to the null.

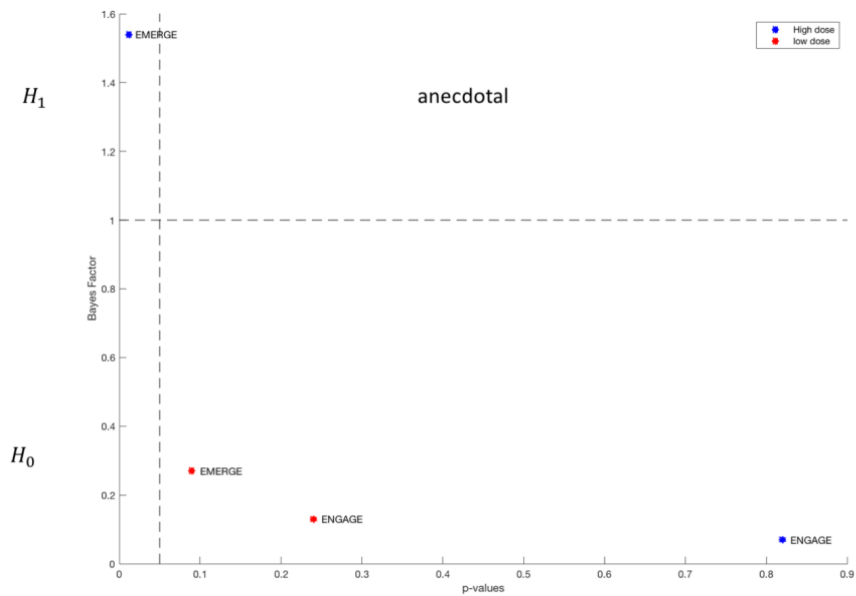

**Fig. 1** Bayes Factor for low and high condition of the EMERGE and ENGAGE trials. Note that only the high dose of the EMERGE trial is in the direction of the alternative hypothesis but at an anecdotal level.

The results show that except the high dose of the EMERGE trial they are all in favor of the null hypothesis. The only data with a value in favor of the alternative hypothesis, drug efficacy, is the high-dose condition in the EMERGE trial. However, the value of the Bayes factor falls within the range of values that are considered anecdotal, i.e. with a low value of evidence. Another way is to transform the Bayes factor to a posterior probability which allows to measure the strength of the evidence provided by the data. The posterior probability can be calculated using the Bayes Factor using the formula:

$$P(H_1|data) = \frac{BF_{10} \times P(H_1)}{BF_{10} \times P(H_1) + P(H_0)}$$

and

$$P(H_0|data) = 1 - P(H_1|data).$$

The posterior probabilities are show in tab.3.

|  | **Low dose** | **High dose** |
|---|---|---|
| EMERGE | $BF_{10} = 0.27$<br>$P(H_1|D) = 21\%$ | $BF_{10} = 1.54$<br>$P(H_1|D) = 60\%$ |
| ENGAGE | $BF_{10} = 0.13$<br>$P(H_1|D) = 11\%$ | $BF_{10} = 0.07$<br>$P(H_1|D) = 9.3\%$ |

**Table 3**. The corresponding posterior probability (expressed in percentages) of the corresponding Bayes Factor for the EMERGE and ENGAGE trials.

The most important results is that the evidence for the high dose in the EMERGE trial is not so important. Only a 60% of evidence the data of the trials support the hypothesis that the high dose drug is effective for the impairment of the Clinical Dementia Rating-Sum of Boxes (CDR-SB).

What is evident from the data is that there is a large difference in the evidence supporting the hypothesis of efficacy of the dosage and of the drug in the two clinical trials. For this reason we decided to calculate the Bayes factor by putting together the two trials and calculating the meta Bayes Factor as described above. The results are shown in fig. 2 and table 4.

|  | **Low dose** | **High dose** |
|---|---|---|
| META-ANALYSIS BAYES FACTOR | $BF_{10} = 0.38$<br>$P(H_11|D)) = 27\%$ | $BF_{10} = 0.29$<br>$P(H_11|D)) = 22\%$ |

**Table 4**. The Bayes Factor and the corresponding posterior probability of evidence in the meta-analysis

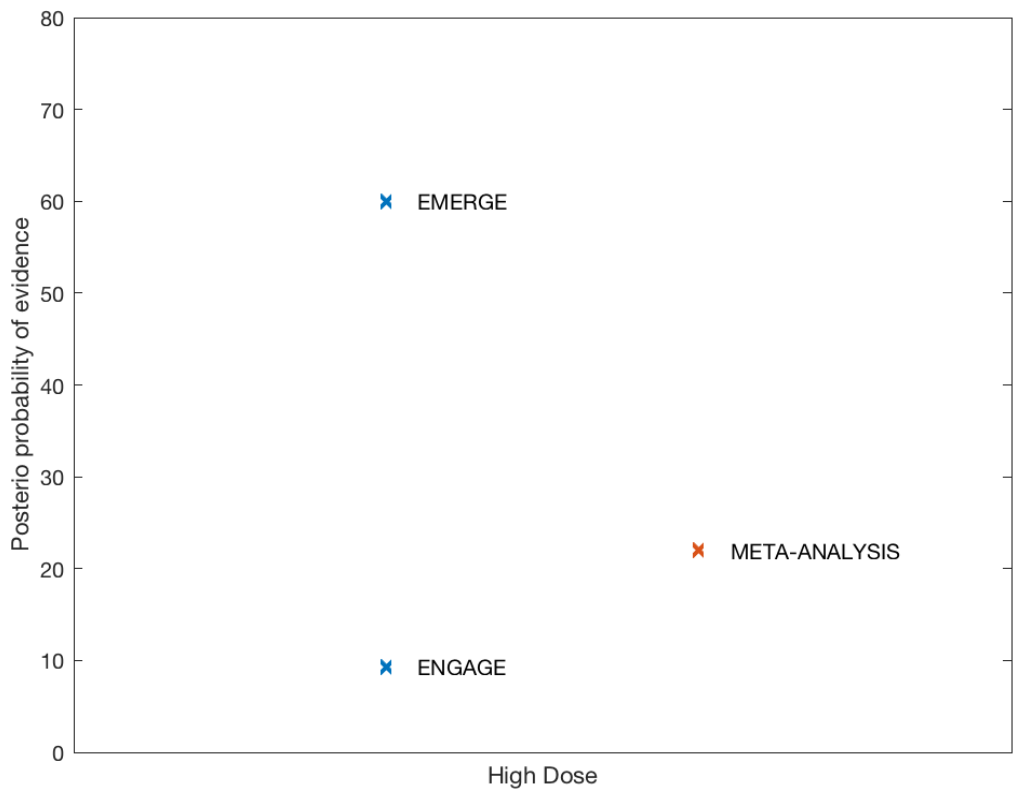

**Fig. 2** The posterior probability of evidence of the efficacy hypothesis for the two trials and combining the results in the meta-analisys.

What is evident from the results is that the evidence of the efficacy of the ADU drugs in the high condition drops drastically showing only a 22% of probability of evidence.

# Conclusion

This study showed that the Bayesian framework provide important and valuable information on the strength of evidence for the efficacy of the High dose Aducanumab drug. In particular, we have shown that in both the separate and the combined conditions the evidence of drug efficacy is very low. these results further highlight the ability of Bayesian methods to provide clearer indications on the hypotheses under study than those obtained in the NHST framework. furthermore, the availability of software such as Jasp or other libraries available in the scientific community have allowed Bayesian method to overcome the difficulty of computation and to obtain results quickly and easily.
These results indicate that it is possible and necessary to adopt Bayesian methods in addition to the NHST framework to render the scientific evidence more meaningful and detailed in particular in the clinical trials.